\documentclass[useAMS,usenatbib]{mn2e}
\usepackage{subfigure}
\usepackage{natbib}
\usepackage{lscape}
\usepackage[dvips]{color}
\usepackage{amssymb}
\usepackage{graphicx}
\usepackage{ulem}

\title{The Ratio of Luminous to Faint Red Sequence Galaxies in X-ray and Optically Selected Low-Redshift Clusters}
\author[Capozzi D. et al. 2009]
 { Diego Capozzi $^{1}$\thanks{E-mail:dc@astro.livjm.ac.uk}, Chris A. Collins $^{1}$ and John P. Stott $^{1}$ \\ \\
 1 - Astrophysics Research Institute, Liverpool John Moores University, Twelve Quays House, Egerton Wharf,\\
  \quad \, Birkenhead, CH41 1LD, UK\\
}
\date{Accepted ;
      Received ;
      in original form }
\pagerange{\pageref{firstpage}--\pageref{lastpage}} \pubyear{2009}

\begin{document}

\maketitle

\label{firstpage}

\begin{abstract}
We study the ratio of luminous-to-faint red sequence galaxies in both optically and X-ray selected galaxy clusters in the poorly studied 
redshift range $0.05\leq z<0.19$. The X-ray selected sample consists of 112 clusters based on the {\it ROSAT} All-Sky Survey, while the 
optical sample consists of 266 clusters from the Sloan Digital Sky Survey. Our results are consistent with the presence of a trend in 
luminous-to-faint ratio with redshift, confirming that downsizing is continuous from high to low redshift.\\ 
After correcting for the variations with redshift using a partial Spearman analysis, we find no significant relationship between 
luminous-to-faint ratio and X-ray luminosity of the host cluster sample, in contrast to recent suggestions. Finally, we investigate the 
stacked colour-magnitude relations of these samples finding no significant differences between the slopes for optically and X-ray selected 
clusters. The colour-magnitude slopes are consistent with the values obtained in similar studies, but not with predictions of theoretical 
models.\\
\end{abstract}


\section{INTRODUCTION}
\label{sec:section1}
It is still unclear how galaxies evolve over the Hubble time and the picture is complicated because galaxy properties also depend on 
environment and mass. In fact, although these dependencies have been extensively studied, it is still an open 
question as to how they are related to the evolution we see. An excellent probe of this evolution is the colour-magnitude relation (CMR) of 
early-type galaxies in clusters and groups, first noted by \citet{Visvanathan-1977} and interpreted as a correlation between galaxy 
mass and mean stellar metallicity \citep{Kodama-Arimoto-1997}. The morphology of the CMR is observed to evolve with redshift 
and its origin can be explained either through concurrent formation of elliptical galaxies in cluster cores in a high redshift monolithic 
collapse \citep{Kodama-Arimoto-1997} or, alternatively, through the hierarchical formation of elliptical galaxies via merging over cosmic 
time \citep{Kauffmann-1998}.\\   
The reliability and the utility of the CMR as a probe of galaxy evolution has also been highlighted by the fact 
that it is an extremely good tool for the identification of galaxy structures like clusters and groups 
(see e.g. \citealp{Gladders-1998, Koester-2007, Swinbank-2007, Capozzi-2009}). Evaluating the relative number of faint and 
luminous red-sequence galaxies (RSGs) in clusters as a function of redshift \citep{De-Lucia-2004,De-Lucia-2007,Stott-2007,Gilbank-Balogh-2008} 
and environment \citep{Tanaka-2005}, is an effective method with which to investigate how galaxies evolve towards the CMR. The tools used to 
quantify the evolution of the red sequence are the faint-end slope of the red-sequence  
luminosity function and the ratio of the number of luminous to faint galaxies ({\rm lum/faint} or, alternatively, giant to dwarf, {\rm g/d}) on the CMR.
There is a current debate concerning which of these two tools is best for undertaking these kinds of studies. For instance, \citet{Andreon-2008} argued
that the use of the faint-end slope is preferable, since it is a measure of the {\rm lum/faint} ratio, it is easier to deal with from a 
statistical point of view and has the advantage of using all the data available. On the other hand, 
\citet{Gilbank-Balogh-2008} pointed out that the dwarf to giant ({\rm d/g}) ratio is just a luminosity function reduced to two bins and 
avoids the complication of having to fit an analytic function, which usually involves degeneracies between the fitted 
parameters.\\

Previous studies carried out to investigate the {\rm lum/faint} ratio as a function of redshift 
have provided conflicting results. Some of them \citep{Barkhouse-2007,Stott-2007,De-Lucia-2007,Gilbank-2008,Hansen-2009} showed results 
consistent with an increasing trend of the {\rm lum/faint} ratio with redshift, while other studies \citep{Tanaka-2005,Andreon-2008} 
indicated that the cluster {\rm lum/faint} ratio may not evolve with redshift. \citet{Tanaka-2005}, using three clusters at different 
redshifts (0, 0.55 and 0.89) found a discordant result only for their $z\sim 0$ cluster, while \citet{Andreon-2008}, studying 28 
clusters at $0.02<z<1.3$ individually, concluded there is no evolution with z. A quite different scenario is the one found by 
\citet{Lu-2009}. In their study of 127 Canada-France-Hawaii Telescope Legacy Survey (CFHTLS) rich clusters with $0.17\leq z \leq 0.36$, in 
comparison with Coma cluster and a sub-sample of 22 groups with $0.08<z<0.09$ taken from the group catalogue by \citet{Yang-2007}, they 
found no strong evolution of the {\rm d/g} ratio (or, similarly, of the faint end of the luminosity function) over the redshift window 
$0.2 \lesssim z \lesssim 0.4$. On the other hand, they also report an increase of a factor of $\sim 3$ from $z\sim 0.2$ to $z \sim 0$.\\  

Several studies have investigated the dependence of the {\rm lum/faint} ratio on the mass of the host systems, by looking for trends with cluster richness, velocity 
dispersion or X-ray luminosity. Unfortunately these studies have also led to contradictory results. 
\citet{Hansen-2009} and \citet{Gilbank-2008} found that the faint-end slope of the cluster red-sequence luminosity function depends on 
cluster richness for $z\lesssim 0.5$, such that low-mass clusters have higher {\rm g/d} ratios than richer systems. 
According to the findings of \citet{De-Lucia-2007}, at intermediate redshifts (0.4-0.8), the {\rm lum/faint} ratios of clusters with velocity dispersion larger than 
$600\ {\rm {km\ s^{-1}}}$ appear to be larger than those measured for clusters at the same redshift but with lower velocity dispersion. 
However, \citet{De-Lucia-2007} pointed out that the error bars and the cluster-to-cluster variations were too large to draw any definitive 
conclusions regarding this point. On the other hand, \citet{Gilbank-Balogh-2008}, using data from three different cluster samples all 
at $z\sim 0.5$ \citep{De-Lucia-2007,Stott-2007,Gilbank-2008}, suggested that the {\rm g/d} ratio is relatively insensitive to mass or 
selection method over the mass range covered by the analysed clusters. They also suggested that the evolution of the cluster {\rm g/d} 
ratio is not due to a systematically changing mass limit with redshift. However, their findings are probably the result of a sample built 
largely from only massive clusters. In fact, only when comparing clusters with large differences in mass (e.g., 
\citealp{De-Lucia-2007,Gilbank-2008} and generally with optically selected samples) is a trend likely to be seen. 
Turning to the studies involving clusters investigated individually, where the cluster-to-cluster scatter is larger
than cases where at least tens of clusters per redshift bin are used, \citet{Koyama-2007} analysed three X-ray selected clusters. They  
studied how the faint-end slope of the luminosity function varied with cluster X-ray luminosity ($L_{\rm X}$), finding quite a steep trend. 
However, this trend, as mentioned by Koyama et al. themselves, is largely based on a sample of inadequate size, given the large intrinsic 
scatter. Finally, \citet{Andreon-2008}, using a sample of 28 X-ray selected clusters, found no correlation between {\rm lum/faint} ratio 
and either $L_{\rm X}$ or velocity dispersion, obtaining the same result utilizing the faint-end slope of the luminosity function.\\

Our work aims to study the evolution of red galaxies in optically and X-ray selected galaxy clusters using data from the Sloan Digital 
Sky Survey Data Release 6 (SDSS DR6), in order to investigate how the {\rm lum/faint} ratio varies between different cluster samples 
(optical and X-ray) and to investigate possible trends with cluster mass and redshift. To parametrize the build up of the CMR, we focus 
our attention on the {\rm lum/faint} ratios and the colour-magnitude relations obtained for two cluster samples, one optically selected 
from SDSS data and the other X-ray selected from the {\it ROSAT} All-Sky Survey data (RASS). We decided to use the {\rm lum/faint} ratio 
as an estimate of the relative number of luminous and faint RSGs, since this is independent on the form of the luminosity function. We 
perform our study of the {\rm lum/faint} ratio in a poorly studied redshift interval ($0.05\leq z<0.19$) (few studies, e.g. 
\citealp{Barkhouse-2009} and \citealp{Lu-2009} have investigated similar redshift windows), as most of the previous studies have focused 
either on the intermediate and high redshift regime ($0.4\lesssim z \lesssim 1.0$) or on Coma-like redshifts ($z\sim 0.02$).\\     

The paper is structured as follows: in Sect. \ref{sec:section2} we describe the cluster samples and the data used in the analysis, while 
Sect. \ref{sec:section3} is dedicated to the data analysis. Sects. \ref{sec:section4} and \ref{sec:section5} are focused on the stacked 
CMRs and on the dependence of the {\rm lum/faint} ratio with redshift, richness, cluster centric distance and $L_{\rm X}$, while in 
Sects. \ref{sec:section6} and \ref{sec:section7} we present and discuss the results. Finally, in Sect. \ref{sec:section8} we draw our 
conclusions.\\
Throughout this paper we make use of magnitudes in the AB photometric system and assume a standard cosmology with 
$H_{0}=70\ {\rm km\ s^{-1}\ Mpc^{-1}}$, $\Omega_{m}=0.3$ and $\Omega_{\lambda}=0.7$.

\begin{figure*}
\begin{center}
\includegraphics[width=0.8\textwidth,angle=0,clip]{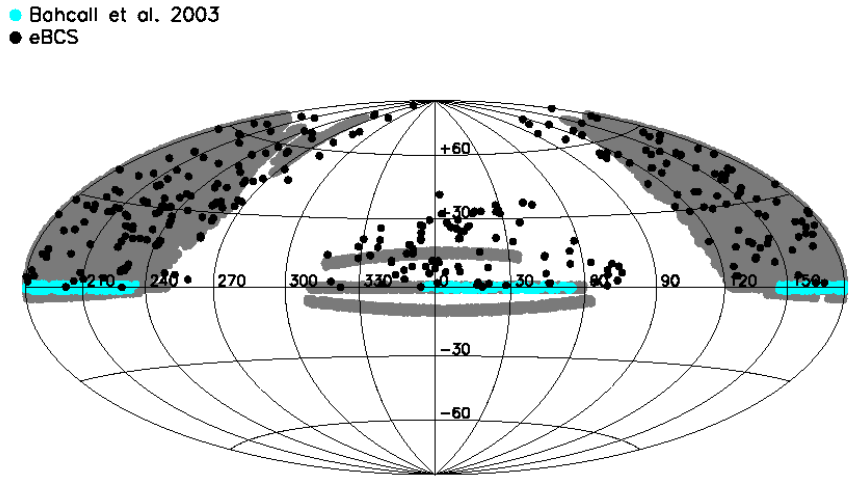} 
\caption{Sky distribution of eBCS \citep{Ebeling-1998, Ebeling-2000} and \citet{Bahcall-2003} cluster samples (black and cyan dots, 
respectively) superimposed on the SDSS DR6 footprint (grey background). An Aitoff projection in equatorial coordinates is used.
The eBCS sample (from which our eBCS subsample is extracted) is made of 310 X-ray selected clusters covering a redshift range of 
$0.02\lesssim z\lesssim 0.42$. The \citet{Bahcall-2003} cluster sample (from which our optical subsamples B and HB are extracted) is 
made of 799 optically selected clusters covering a redshift range of $0.05<z<0.3$.}
\label{fig:Fig1}
\end{center}
\end{figure*}

\section{SAMPLE SELECTION \& DATA}
\label{sec:section2}
\begin{figure}
\begin{center}
\includegraphics[width=0.4\textwidth,angle=0,clip]{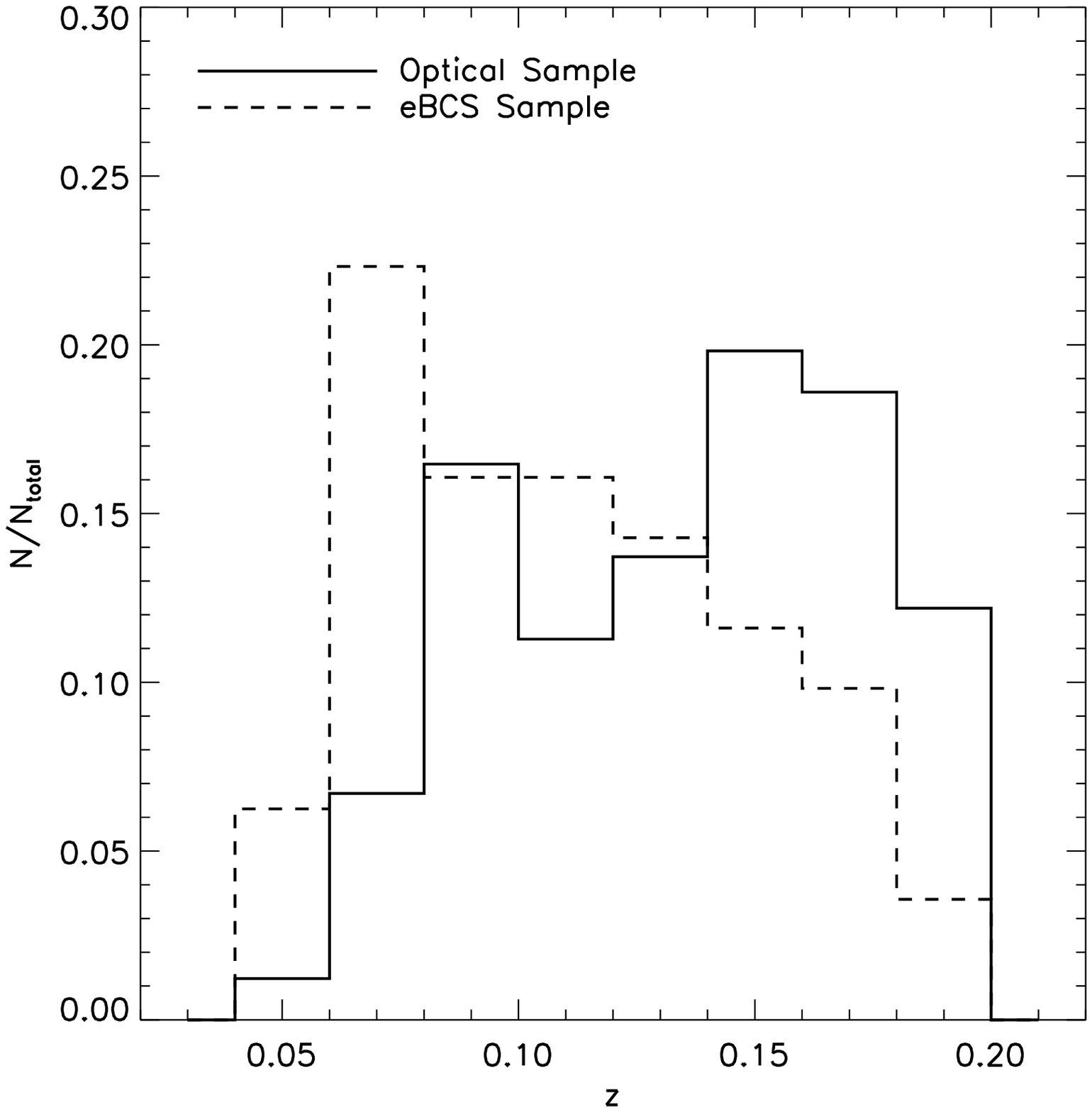} 
\caption{Redshift distribution of the analysed clusters taken from eBCS \citep{Ebeling-1998,Ebeling-2000} and \citet{Bahcall-2003} cluster 
samples.}
\label{fig:Fig2}
\end{center}
\end{figure}

To perform our study, we utilize two cluster samples composed of X-ray and optically selected systems, respectively; their sky distribution, 
superimposed on the SDSS DR6 footprint, and the redshift distribution are shown in  Figs. \ref{fig:Fig1} and  \ref{fig:Fig2}, respectively.\\

\noindent The X-ray selected cluster sample contains 112 clusters with $0.05\leq z<0.19$ falling into the SDSS DR6 footprint, included in the
homogeneously selected extended Brightest Cluster Sample (eBCS, \citealp{Ebeling-1998,Ebeling-2000}). This sample is made up of two cluster 
catalogues both selected from the RASS data in the northern hemisphere ($\delta \geq 0\ {\rm deg}$) and at high 
Galactic latitudes ($|b|\geq 20\ {\rm deg}$): (i) a 90 per cent flux-complete sample (called the {\it ROSAT} Brightest Cluster Sample, 
BCS) consisting of the 201 X-ray brightest clusters in the RASS data, with measured redshifts $z \leq 0.3$ and fluxes higher than 
$4.4\times 10^{-12}\ {\rm erg\ cm^{-2}\ s^{-1}}$ in the $0.1-2.4\ {\rm keV}$ band; (ii) a low-flux extension of the BCS comprising 107 
X-ray clusters of galaxies with measured redshifts $z\leq 0.42$ and total fluxes between 
$2.8\times 10^{-12}$ and $4.4\times 10^{-12}\ {\rm erg\ cm^{-2}\ s^{-1}}$ in the $0.1-2.4\ {\rm keV}$ band (the latter value being the flux 
limit of the original BCS).\\
\noindent X-ray fluxes have been computed using an algorithm tailored for the detection and characterization of X-ray emission from galaxy 
clusters (\citealp{Ebeling-2000}) and the fluxes are accurate to better than 15 per cent ($1\sigma$ error). The nominal completeness of the 
eBCS sample, defined with respect to a power-law fit to the bright end of the BCS $\log{N}-\log{S}$ distribution (see Fig. 2 in 
\citealp{Ebeling-2000}), is 75 per cent, compared with 90 per cent for the high-flux BCS.\\ 
We use the fluxes published in \citet{Ebeling-1998,Ebeling-2000} to calculate cluster X-ray luminosities according to the 
cosmological model used in this work.\\

\begin{table*}
%\begin{tiny}
\begin{center}
\begin{tabular}{lclcllcllcl}
\hline
{\bf Sample} &  {\bf slope}	                 &  $\mathbf{z\ bin_{1}}$ & $\mathbf{\rm{\bf lum_{1}}} \| \mathbf{\rm{\bf faint_{1}}}$ & $\mathbf{\rm{\bf lum/faint_{1}}}$ &  $\mathbf{z\ bin_{2}}$ & $\mathbf{\rm{\bf lum_{2}}} \| \mathbf{\rm{\bf faint_{2}}}$ & $\mathbf{\rm{\bf lum/faint_{2}}}$            \\  			  
\hline  
  eBCS	    &   $-0.036^{+0.001}_{-0.001}$    &   	 0.08 	         &                         $25\|45\ $ 						       &	$0.47\pm0.01$  					    &	0.15  					  &  	                      $34\|56\ $ 							   &   $0.60\pm0.02$ 						   		  \\  					 
                                                   					   	   		   					   																																																															
  B	       &   $-0.035^{+0.003}_{-0.002}$    &   	 0.09 	         &                         $12\|22\ $ 						       &	$0.46\pm0.02$  					    &	0.16  					  &  	                      $13\|22\ $ 							   &   $0.52\pm0.02$ 						   		  \\  					 
                                                   							   		   					   																																																															
  HB	       &   $-0.032^{+0.002}_{-0.003}$    &   	 0.08 	         &                         $13\|25\ $ 						       &	$0.41\pm0.02$  					    &	0.15  					  &  	                      $14\|25\ $ 							   &   $0.43\pm0.02$ 						   		  \\  					 
\hline
\end{tabular}
\caption{Values of stacked CMR slope (see Fig. \ref{fig:Fig3}), average number of luminous and faint RSGs per 
redshift bin and background corrected {\rm lum/faint} ratio (see Fig. \ref{fig:Fig4}) per redshift bin for the analysed cluster samples.  
The {\rm lum/faint} ratio calculated in the background fields is always comprised between 0.19 and 0.22. The error values of the slopes 
correspond to the 95 per cent confidence interval as measured on the bootstrap distributions.}
\label{tab:Tab1}							    
\end{center}
%\end{tiny}
\end{table*}

\noindent The optically selected cluster sample is the one presented by \citet{Bahcall-2003} containing 799 clusters of galaxies in the 
redshift range $z=0.05-0.3$ and selected from $\sim 400\ {\rm deg^{2}}$ of early SDSS 
commissioning data along the celestial equator. Clusters have been found through the application of two independent identification
algorithms: a colour-magnitude red-sequence maxBCG technique (\citealp{Koester-2007}) and a hybrid matched filter 
method (hereafter HMF, \citealp{Kim-2002}). These two algorithms focus on different properties of galaxy clusters. The maxBCG uses a 
brightest cluster galaxy (BCG) likelihood based on luminosity and colour applied to each SDSS galaxy weighted by the number of nearby 
galaxies located within the CMR appropriate to E and SO galaxies. The algorithm therefore selects clusters dominated by bright red 
galaxies. In contrast the HMF uses a model Plummer density profile and a Schechter luminosity function \citep{Schechter-1976} with 
typical parameters observed for galaxy clusters and is sensitive to the galaxy population fainter than $L^{\ast}$. The use of both maxBCG and 
HMF selected clusters enables us to include determinations of the CMR for representative cluster selection algorithms based on galaxy 
colour and density profile. The optical sample contains clusters with richness $\Lambda \geq 40$ (HMF richness) and $N_{gal} \geq 13$ 
(maxBCG richness), which translates into a mean cluster velocity dispersion of $\sigma_{r}\gtrsim 400\ {\rm {km s^{-1}}}$.\\ 
 We refer to the original papers for a detailed description of these two algorithms.\\ 

For our analysis we utilize the SDSS DR6 public archive, which covers $9,583\ {\rm deg^{2}}$ of the celestial 
sphere in 5 bands ({\it ugriz})\footnote{A detailed description of the survey can be found at 
http://www.sdss.org/}.

\section{Analysis}
\label{sec:section3}
For the clusters in all samples we extract photometric data from the SDSS DR6 data base. We exclude clusters located at the borders of the 
DR6 footprint and select only clusters in the redshift range $z=0.05-0.19$, where the highest z is chosen to remain within the magnitude 
completeness level of SDSS. So, we are finally left with 112 (eBCS sample) and 266 (optical sample) clusters. 
We split the optical sample into two subsamples according to the selection method used: B subsample (181 clusters) containing 
maxBCG clusters; HB subsample (156 clusters) made of HMF clusters. These two subsamples partially overlap (71 B clusters are 
included in the HB subsample).\\ 
In our analysis we use the dereddened model magnitudes from SDSS, corrected for AB offsets. To perform 
k corrections we always utilize the software developed by \citet{Blanton-2007} for creating template sets based on stellar population synthesis 
models from a set of heterogeneous photometric and spectroscopic galaxy data. The technique, suitable for estimating k corrections for 
ultraviolet, optical and near infrared observations in the redshift range $0<z<1.5$, is based on the non-negative matrix factorization 
method, which is akin to principal component analysis. The templates are fitted to data from Galaxy Evolution Explorer (GALEX), 
SDSS, Two-Micron All Sky Survey (2MASS), the Deep Extragalactic Evolutionary Probe (DEEP) and the Great Observatories Deep Survey
(GOODS). We refer to the original paper for further details. We always use the values of \citet{Poggianti-1997} to correct for passive
evolution.

\begin{figure*}
\begin{center}
\includegraphics[width=0.48\textwidth,angle=0,clip]{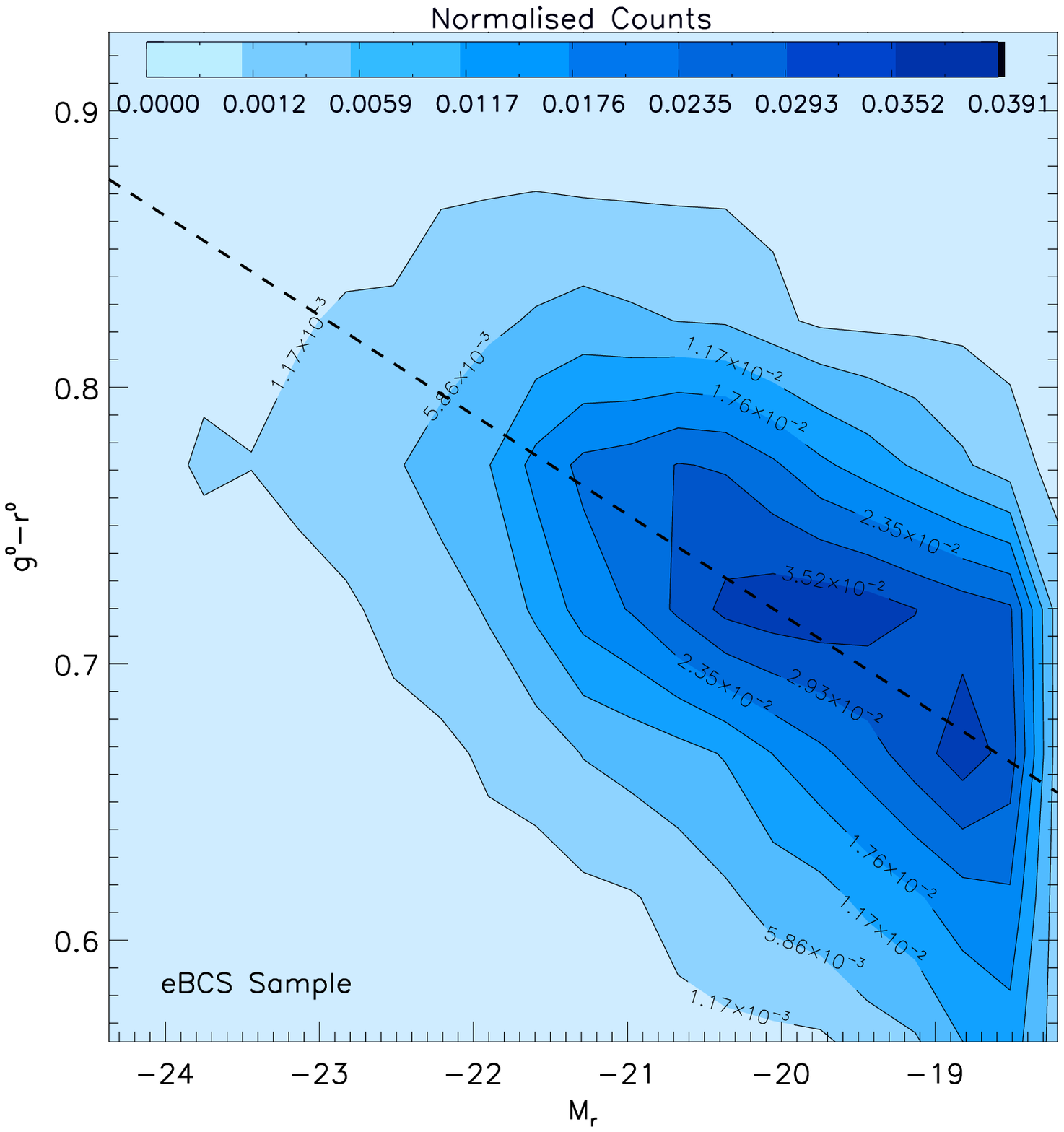} 
\includegraphics[width=0.48\textwidth,angle=0,clip]{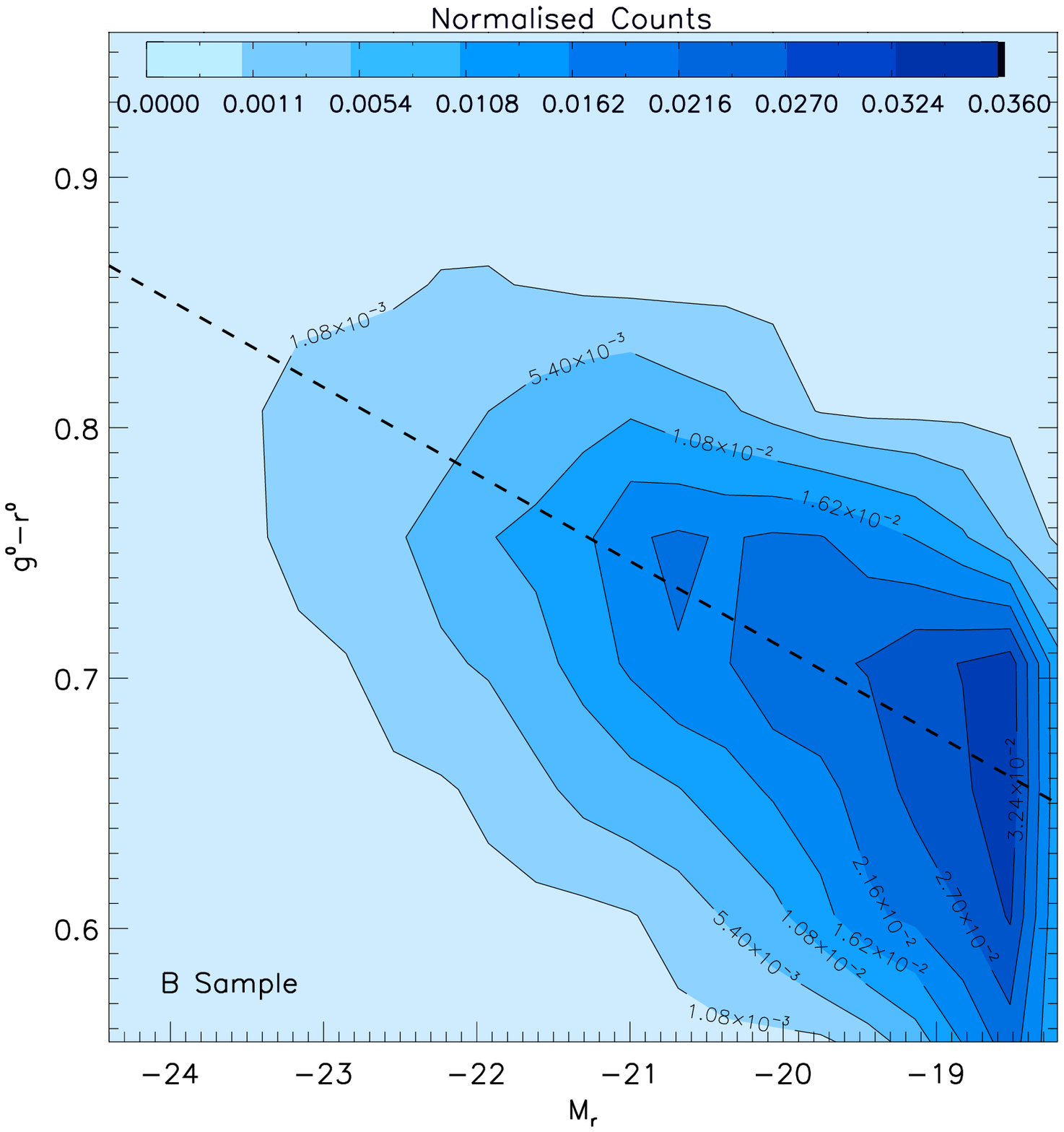} 
\includegraphics[width=0.48\textwidth,angle=0,clip]{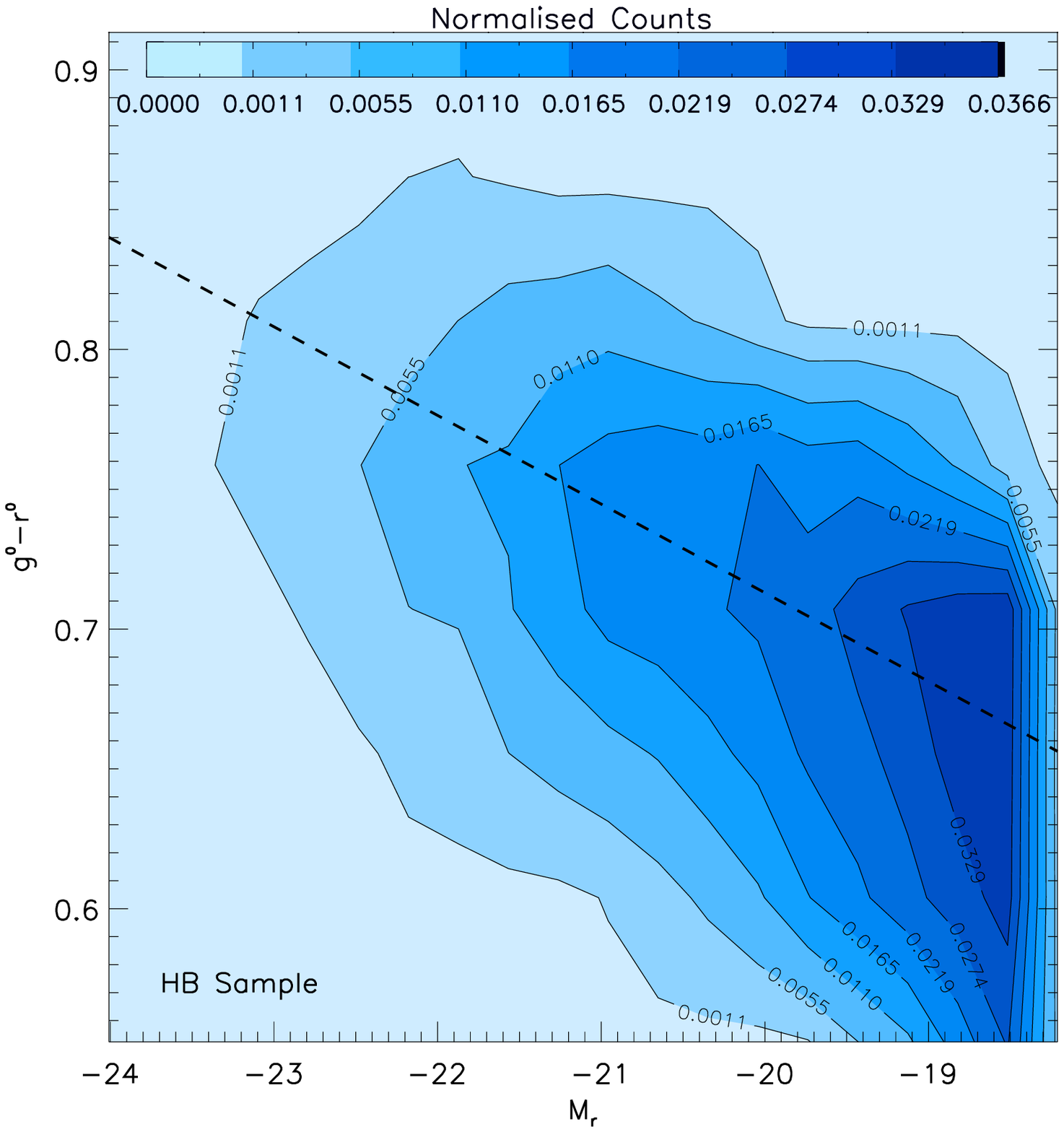} 
\caption{Density maps of the stacked {\it g-r} vs. $M_{\rm r}$ colour-magnitude diagrams for eBCS (upper left panel), B 
(upper right panel) and HB (lower panel) samples. All counts per bin (of sides 0.3 in magnitude and 0.05 in colour) are normalized 
to the value of the total counts. Isodensity contour levels are superimposed and colour coded (the lighter the colour, the less 
counts) according to the colour bar in the plots. The best-fitting CMR (dashed line) is superimposed. }
\label{fig:Fig3}
\end{center}
\end{figure*}

\section{Colour-Magnitude Relation}
\label{sec:section4}
We obtain the stacked {\it g-r} vs. {\it r} CMR for all the analysed samples (Fig. \ref{fig:Fig3}), using red 
galaxies within $1\ {\rm Mpc}$ from the cluster centroids in the cluster's rest-frame.
We first apply a k correction and calculate distances assuming all galaxies are at their cluster's mean
redshift. Then we perform a biweight fit \citep{Beers-1990} on the stacked colour-magnitude diagram using all galaxies with
$M_{r}\leq -21$ (this limit is chosen in order to avoid excess noise in the CMR). From this, we obtain an estimate of the 
slope and the zero-point of the stacked CMR. The biweight fit is performed iteratively using only those galaxies located within 
$0.2\ {\rm mag}$ of the previous CMR best-fitting line.\\
To correct for passive evolution, we use the trends given by \cite{Poggianti-1997} for the redshift range used in this study 
($0.05\leq z<0.19$). Over this interval, the correction is virtually linear and is almost independent of the morphological type. The 
correction is calculated by means of a linear interpolation between the values given by Poggianti and converted to the {\it r} band. Only 
galaxies within $\pm 0.1\ {\rm mag}$ of the CMR best-fitting line are corrected for passive-evolution, to minimize contamination by blue 
galaxies. Subsequently, we perform the last biweight fit using only passive-evolution corrected galaxies (Fig. \ref{fig:Fig3}) to obtain a 
final CMR best-fitting line. To measure the accuracy of the best-fitting CMR parameters, we adopt a bootstrap technique and resample, with 
replacement, the clusters constituting the stacked colour-magnitude distribution 1000 times. By carrying out the same biweight fit 
we derive the marginalized 2-$\sigma$ confidence levels on the measured parameters from their bootstrap distributions.
We perform this analysis on all the cluster samples (eBCS, B, BH). The final results for the CMR slope are reported in Tab. \ref{tab:Tab1}.

\section{Luminous to Faint Ratios} 
\label{sec:section5}
Following the approach used by \citet{De-Lucia-2007}, we split the galaxies on the red sequence into luminous (or giant) and faint 
(or dwarf) galaxies for each cluster.  \citet{De-Lucia-2007} classify all galaxies having $M_{\rm {V}}<-20$ as luminous and those galaxies 
with $-20<M_{\rm {V}}<-18.2$ as faint; these limits are valid for galaxies whose magnitudes have been corrected for passive evolution to 
z = 0. The Johnson {\it V} band magnitude can be computed from SDSS photometry using the following relation:

\begin{equation}
V = r + 0.44*(g-r) - 0.02,
\end{equation}  

\noindent which has an accuracy better than 0.05 mag \citep{Fukugita-1996}. We use this transformation to convert the De Lucia et al. 
magnitude limits from {\it V} to {\it r} band, utilizing the colour {$ g-r=0.77$} computed by \citet{Fukugita-1995} for an elliptical 
galaxy at z = 0. After this transformation, we obtain a faint and a bright absolute magnitude limit of $M_{\rm {r}}=-18.52$ and 
$M_{\rm {r}}=-20.32$ respectively.
To calculate the number of faint and luminous galaxies on the colour-magnitude relation of our clusters, we perform a biweight fit 
\citep{Beers-1990} on the apparent colour-magnitude diagram ({\it g-r} vs. {\it r}) of each cluster to determine an individual 
best-fitting CMR, using only galaxies within $1\ {\rm Mpc}$ of the cluster centroids.\\
\noindent Hereafter RSGs refer to galaxies within $\pm 0.3\ {\rm mag}$ of each individual CMR best-fitting line. 
To determine the number of faint and luminous RSGs, we need to transform the faint and luminous absolute magnitude limits previously 
discussed, into apparent magnitudes (hereafter $m_{\rm {faint}}$ and $m_{\rm {lum}}$). We perform this transformation using a mean 
k correction value calculated by averaging all the galaxies within $1\ {\rm Mpc}$ of the cluster centroids and applying a 
passive-evolution correction inferred as in Sect. \ref{sec:section4}. At this point, we determine the number of faint 
($m_{\rm {lum}}\leq m_{\rm {r}}\leq m_{\rm {faint}}$) and luminous ($m_{\rm {r}}<m_{\rm {lum}}$) RSGs per cluster within 
$0.75\ {\rm Mpc}$ of their centroids.\\ 
Note that the choice of a radial distance of $0.75\ {\rm Mpc}$ is taken in order to be consistent with the
analysis performed by \citet{De-Lucia-2007}.\\
The canonical photometric completeness limit of SDSS is defined as the 95 per cent detection repeatability for point sources 
({\it g}=22.2; {\it r}=22.2, as reported on SDSS website). However, the extended profile of early-type galaxies outside the point spread function may result 
in incompleteness at a brighter magnitude limit. In order to estimate the completeness level, we use \citet{Cross-2004}, who studied
the completeness of SDSS Early Data Release by comparing galaxy number counts as a function of surface brightness, using the overlapping 
region of the deeper Millennium Galaxy Catalogue. For galaxies with $\mu_{50_{\rm r}}\lesssim 24.6$ 
the completeness is $>95$ per cent, while for galaxies with $24.6 \lesssim \mu_{50_{\rm r}} \lesssim 25.6$ it reduces to $\sim 90$ per cent.
Using the same definition of surface brightness of \citet{Cross-2004}, for our data in the highest redshift interval ($0.15<z<0.19$) this 
translates to losing a fraction of faint galaxies with $24.6 \lesssim \mu_{50_{\rm r}} \lesssim 25.6$ corresponding to the 0.5 per cent of 
the total number of faint galaxies. Therefore, we think the effect of incompleteness is negligible and does not affect our results.\\ 

\begin{figure*}
\begin{center}
\includegraphics[width=0.6\textwidth,angle=0,clip]{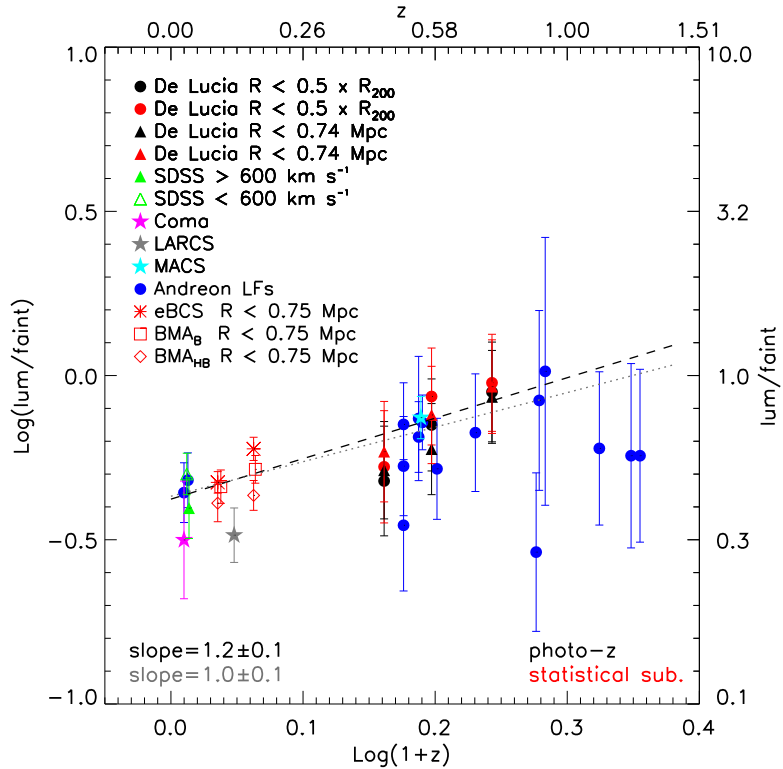} 
\caption{Luminous-to-faint ratios as function of redshift [$\log({\rm lum/faint})$ vs. $\log(1+z)$] obtained using the mean background 
method. \citet{De-Lucia-2007} values: $R< 0.5 \times R_{200}$ (black and red plain dots), $R< 0.74\ {\rm {Mpc}}$ (black and red plain 
triangles), SDSS clusters with $\sigma_{\rm {r}}>600\ {\rm {km\ s^{-1}}}$ (plain green triangle), SDSS clusters with 
$\sigma_{\rm {r}}<600\ {\rm {km\ s^{-1}}}$ (open green triangle), Coma cluster (magenta plain star). \citet{Stott-2007} values: LARCS 
(grey plain star), MACS (cyan plain star). Blue full dots represents \citet{Andreon-2008} values. Red asterisks, crosses and plain squares
represent our values for eBCS, HB and B samples respectively. Red symbols have been inferred through a statistical approach, while black 
ones have been derived through the application of photometric redshifts. The best-fitting line obtained through a weighted fit carried out 
excluding values inferred through individual cluster measurements is plotted (dashed black line). The dotted grey line is obtained when 
the single cluster measurements are included (in both cases clusters from \citet{Andreon-2008} with $z>1$ are excluded, see the text). The 
errors on the slopes are evaluated through a jackknife technique.} 
\label{fig:Fig4}
\end{center}
\end{figure*}

\subsection{Background Subtraction} 
\label{subsec:section5.1} 
We utilize two approaches to evaluate the numbers of background galaxies contaminating the estimates of faint and luminous RSGs. 
The first one makes use of 17 control fields \citep{De-Filippis-2009}, randomly selected 
within the SDSS footprint, within which, after an a posteriori check, no large local structures are found (Tab. \ref{tab:Tab2}, 
total area of $16.36\ {\rm deg^{2}}$). We determine the number of faint and luminous galaxies within $\pm 0.3\ {\rm mag}$ from the 
cluster CMR in each background region and after normalising them for the area, we calculate a weighted mean of the obtained values.
The second approach, utilizes a local background, i.e. an annular region between 2 and 3 {\rm Mpc}. The numbers of 
faint and luminous galaxies are determined in the same way as the mean background approach.
Comparing the final background subtracted distributions of {\rm lum/faint} ratios for the two methods we obtain very similar results 
within $1 \sigma \sim 0.03$ and therefore in what follows we present the results only for the mean background method.

\begin{table}
\begin{center}
\begin{tabular}{|l|c|c|c|}
\hline
  \multicolumn{1}{|c|}{{\bf bgID}} &
  \multicolumn{1}{c|}{{\bf RA} (degrees)} &
  \multicolumn{1}{c|}{{\bf Dec} (degrees)} &
  \multicolumn{1}{c|}{{\bf radius} (arcmin)} \\
\hline
  BG00001 & 180.0 & 54.0 & 21.0\\
  BG00002 & 170.0 & 53.0 & 35.0\\
  BG00003 & 120.0 & 12.0 & 60.0\\
  BG00004 & 120.0 & 14.0 & 22.0\\
  BG00005 & 120.0 & 20.0 & 27.0\\
  BG00006 & 120.0 & 22.0 & 19.0\\
  BG00007 & 120.0 & 26.0 & 23.0\\
  BG00008 & 125.0 & 24.0 & 21.0\\
  BG00009 & 125.0 & 26.0 & 21.0\\
  BG00010 & 130.0 & 24.0 & 55.0\\
  BG00011 & 130.0 & 20.0 & 20.0\\
  BG00012 & 140.0 & 22.0 & 47.0\\
  BG00013 & 140.0 & 20.0 & 24.0\\
  BG00014 & 240.0 & 42.0 & 23.0\\
  BG00015 & 230.0 & 40.0 & 37.0\\
  BG00016 & 230.0 & 52.0 & 40.0\\
  BG00017 & 315.0 & 0.0 & 28.0\\
\hline\end{tabular}
\caption{Control fields used for the background subtraction \citep{De-Filippis-2009}.}
\label{tab:Tab2}							    
\end{center}
\end{table}

\subsection{Dependence on Redshift, Radius, Richness and $\mathbf{L_{X}}$}
\label{subsec:section5.2}
\noindent To study the relationship between {\rm lum/faint} ratios and redshift we subdivide clusters in two redshift bins. In each of 
these bins, we then calculate a weighted mean of their ratios (Tab. \ref{tab:Tab1}). A potential problem is that, below $z\sim 0.1$, 
the $4000\ {\rm \AA}$ break is beginning to slip to the extreme blue edge of the {\it g} filter (at our minimum redshift of 0.05, the 
Balmer break falls at $\sim 4200\ {\rm \AA}$ whereas the {\it g} filter's waveband starts approximately at $4000\ {\rm \AA}$), potentially 
biasing our estimates of the ratios in the lowest redshift bin. We test this possibility by recalculating the ratios in this redshift bin 
using the {\it u-r} filter combination. For the eBCS sample, for instance, we obtain a value of ${{\rm lum/faint}_{1}}=0.46\pm 0.02$, very 
similar to the value obtained for the same sample using the {\it g-r} colour, which is ${{\rm lum/faint}_{1}}=0.47\pm 0.01$.\\
We compare our results (Fig. \ref{fig:Fig4}) together with other literature estimates over the redshift range $0.02<z<1.3$.\\
\noindent We also investigate the presence of a trend of the {\rm lum/faint} ratio with cluster-centric distance, since, as 
highlighted recently by \citet{Barkhouse-2009}, the use of a fixed physical aperture, instead of one scaled to the cluster's virial radius,
may cause the {\rm lum/faint} ratios to be over estimated for more massive clusters. However, our adopted radius of $0.75\ {\rm Mpc}$ 
covers a fraction of the virial radius in $0.3\lesssim \frac{0.75\ {\rm Mpc}}{r_{\rm vir}} \lesssim 0.5$ for the eBCS sample, 
over which the {\rm lum/faint} ratio should not evolve significantly (Fig. 2 of \citealp{Barkhouse-2009}). For similar reasons, we do not 
expect these issues to significantly affect the optical sample either. However, since our methodology of estimating the {\rm lum/faint} 
ratio and the one used by \citet{Barkhouse-2009} might not be directly interchangeable, we further investigate its trend with 
cluster-centric distance. In order to highlight differences, we test the change of the {\rm lum/faint} ratio with radius by recalculating 
its values using an aperture of $1.5\ {\rm Mpc}$ (corresponding to $0.6\lesssim \frac{1.5}{r_{\rm vir}} \lesssim 1$ for the eBCS sample). 
We find no significant differences in the values of the {\rm lum/faint} ratio (e.g., for the B sample, in order of increasing z: 0.47, 
0.52).\\
\noindent In addition, we study the relation between {\rm lum/faint} ratio and cluster richness looking for correlations between 
{\rm lum/faint} and {\rm lum} RSGs and between {\rm lum} and {\rm faint} RSGs. For this purpose we use both the full scatter plots and 
the ones obtained for each redshift bin. When performing Spearman's rank correlation tests on these plots, no significant correlation 
is found (r values are always about 0.4 for lum/faint vs. lum correlation and 0.5 for lum vs. faint correlation).\\
\noindent Finally, in order to further investigate the dependence of the {\rm lum/faint} ratio on cluster mass, we study it as 
a function of $L_{\rm X}$ for the eBCS sample. We subdivide this sample according to z and $L_{\rm X}$ in order to obtain 
approximately equally populated volume limited bins (Fig. \ref{fig:Fig5}). We then analyse the ratio as a function of $L_{\rm X}$ in each 
redshift bin. Our results are shown in Fig. \ref{fig:Fig6}, where a mass scale is also shown, inferred by using the $M_{\rm 200}-L_{\rm X}$ 
relation of \citet{Popesso-2005}.

\section{Results}
\label{sec:section6}
Our study of the {\rm lum/faint} ratio yields statistical results for all three of our samples which are consistent
with those found by \citet{De-Lucia-2007} for SDSS clusters (Fig. \ref{fig:Fig4}). We test the correlation of {\rm lum/faint} ratio
with z [in terms of $\log({\rm lum/faint})$ and $\log(1+z)$] by performing a Spearman rank correlation test on only the values based on 
cluster samples (\citealp{De-Lucia-2007,Stott-2007} and this work). We find a rank correlation coefficient of $0.89$ with a two sided significance of its deviation from zero of 
$4\times 10^{-8}$. We also perform a weighted fit on the same points plotted in Fig. \ref{fig:Fig4} (dashed line), obtaining a slope of 
$1.2\pm 0.1$. The error is obtained through a jackknife technique, in order to probe the stability of the trend. In determining the best 
fit, we prefer to exclude the {\rm lum/faint} ratio values obtained for individual clusters (the values of \citealp{Andreon-2008} and of 
\citealp{De-Lucia-2007} for the Coma cluster), because of the scatter that individual clusters may introduce. In addition, the highest 
redshift clusters ($z>1$) included in the sample by \citet{Andreon-2008} have been observed with filters that do not bracket the 
$4000\ {\rm {\AA}}$ break as adequately as the remaining lower redshift clusters, leading to a potential source of contamination of blue 
galaxies on the red sequence. In fact, excluding them, 12 of the 13 remaining points seem to be in agreement with an increasing trend of 
the {\rm lum/faint} ratio with z. Including the values of \citet{Andreon-2008} (clusters at $z<1$) and Coma, the trend with z is shallower 
but still present (slope value$=1.1\pm 0.1$, dotted line in Fig. \ref{fig:Fig4}).
Accordingly, our low-z samples support the general trend interpreted as evidence of downsizing in the CMR, i.e. relatively more luminous 
galaxies at high z compared to their lower luminous counterparts. The presence of downsizing is in accordance with 
\citet{Barkhouse-2007,Stott-2007,De-Lucia-2007,Gilbank-2008,Hansen-2009,Lu-2009}, but is in contrast to the results of \citet{Tanaka-2005} and 
\citet{Andreon-2008}. The \citet{Lu-2009} analysis was carried out in a similar redshift window to our work 
here. They estimated the {\rm d/g} ratio using a model CMR calibrated on Coma. This technique gives {\rm d/g} ratios $40-60$ per cent 
lower than our values. However, these values can be reconciled as this difference reduces to 25 per cent when a similar colour cut 
is used to the one presented here.\\  

\begin{figure}
\begin{center}
\includegraphics[width=0.45\textwidth,angle=0,clip]{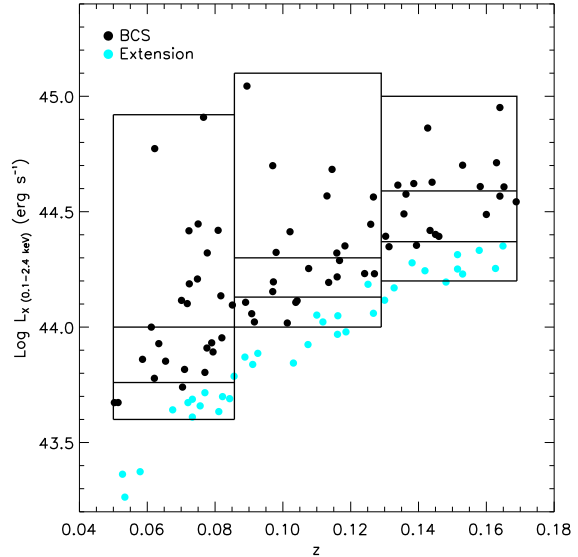} 
\caption{eBCS clusters. Cluster binning in equally populated z and $L_{\rm X}$ bins. Black full dots represent BCS clusters 
\citep{Ebeling-1998} while cyan ones stand for extension clusters \citep{Ebeling-2000}} 
\label{fig:Fig5}
\end{center}
\end{figure}

\begin{figure}
\begin{center}
\includegraphics[width=0.45\textwidth,angle=0,clip]{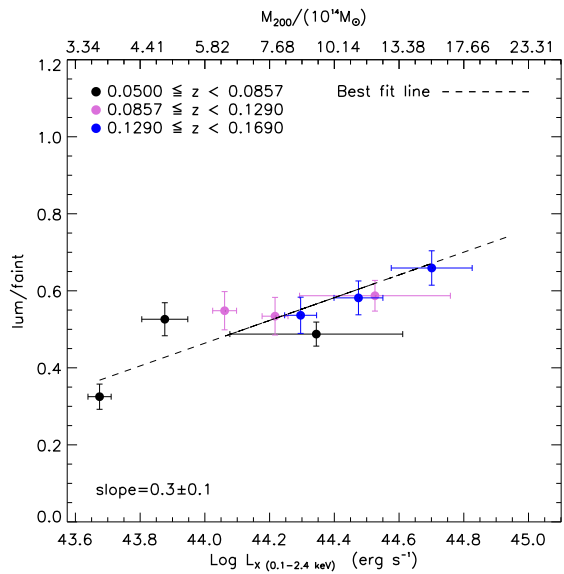} 
\caption{eBCS clusters. Luminous to faint ratios as function of $\log(L_{\rm X})$. The values reported in the plot are the average values 
of the {\rm lum/faint} ratios and of $\log(L_{\rm X})$ obtained in the z and $L_{\rm X}$ bins shown in Fig. \ref{fig:Fig5}. A mass scale 
is also shown, inferred using the $M_{\rm 200}-L_{\rm X}$ relation of \citet{Popesso-2005}. The best-fitting line (dashed line), whose slope 
is displayed at the lower left corner of this plot, is reported. The error on the slope is evaluated through a jackknife technique.} 
\label{fig:Fig6}
\end{center}
\end{figure}

Turning to the {\rm lum/faint} ratio as a function of $L_{\rm X}$, we initially find a possible trend, according to which the 
{\rm lum/faint} ratio increases with $L_{\rm X}$ (Fig. \ref{fig:Fig6}), and, as a consequence, with cluster mass. However, the trend seen 
between the {\rm lum/faint} ratio and $L_{\rm X}$ could be the result of the correlations between $L_{\rm X}$ and z and between 
{\rm lum/faint} ratio and z (see Figs. \ref{fig:Fig4} and \ref{fig:Fig5}). To test this possibility, we compute the partial Spearman 
correlation coefficients using the unbinned data. These coefficients are of the form 
$r_{_{\rm AB,C}}$, where

\begin{equation}
r_{_{\rm AB,C}}=\frac{r_{_{\rm AB}}-r_{_{\rm AC}}r_{_{\rm BC}}}{[(1-{r_{_{\rm AC}}}^{2})(1-{r_{_{\rm BC}}}^{2})]^{1/2}},
\end{equation}  

\noindent and $r_{_{\rm AB}}$, etc., are the Spearman's rank correlation coefficients. Assuming that 
$r_{_{\rm AB,C}}[(N-3)/(1-{r_{_{\rm AB,C}}}^{1/2})]$ (where N represents the number of points) is distributed like a Student's t-statistic,
the significance of the partial rank correlation coefficients may be calculated  to compute the probability, $P(r_{_{\rm AB,C}})$, of 
obtaining a partial rank correlation coefficient with absolute value as large as $|r_{_{\rm AB}}|$, or larger, under the null hypothesis that the 
correlation between A and B results solely from correlations between A and C and between B and C. The results of this test are reported in 
Tab. \ref{tab:Tab3}, where we see that the correlation between {\rm lum/faint} ratio and $L_{\rm X}$ at fixed z is negligible 
($r_{_{\rm AB,C}}=0.02$) and is consistent with arising only from the correlations between $L_{\rm X}$ and z and between {\rm lum/faint} 
ratio and z. This is in accordance with the results of \citet{Gilbank-Balogh-2008} and \citet{Andreon-2008}. However it must be said 
that, similarly to \citet{Gilbank-Balogh-2008}, the lack of a trend between {\rm lum/faint} ratio and $L_{\rm X}$ could be due the small 
range in cluster mass examined (predominantly high mass systems).\\

The biweight fit performed on the stacked colour-magnitude diagrams of the cluster samples studied in this work,
produces estimates of the CMR slope consistent among themselves within $2 \sigma$. 
The values shown are about $-0.034$, in accordance with 
other studies based on observations (e.g. \citealp{Stott-2009}).
Despite this, as reported in similar studies, a discrepancy is seen when 
the CMR slopes are compared with the findings inferred through theoretical models. In fact, our values are not 
consistent with any of those obtained through the model by T. Kodama (see \citealp{Kodama-Arimoto-1997} for the
description of the model) in the SDSS bands for several galaxy formation redshifts. This is probably due to the fact that 
our slope values are obtained for stacked CMRs, containing several and possibly diverse clusters, while this model is calibrated 
to the CMR of the Coma cluster.\\ 

\begin{table}
\begin{center}
\begin{tabular}{ccc}
\hline
$\mathbf r_{_{\rm AB,C}}$  & $\mathbf r_{_{\rm AC,B}}$ & $\mathbf r_{_{\rm BC,A}}$  \\ 
\hline  
            0.02 ($\sim 45\%$)    &            0.15 ($\sim 2.5 \%$)    &              0.66 ($<0.05\%$)          \\
\hline
\end{tabular}
\caption{Results of partial Spearman rank correlation tests for eBCS cluster sample. The tests are performed for {\rm lum/faint} ratios 
(A), $\log(L_{X})$ (B) and z (C). In brackets is reported the probability of obtaining a partial rank correlation coefficient with absolute 
value as large as, for instance, $|r_{_{\rm AB}}|$, or larger, under the null hypothesis that the correlation between A and B results 
solely from correlations between A and C and between B and C.}
\label{tab:Tab3} 							    
\end{center}
\end{table}

\section{Discussion}
\label{sec:section7}

Tab. 1 shows that the {\rm lum/faint} ratios between optical and X-ray clusters vary by as much as $30$ per cent within a single redshift bin, 
which, on its own, goes some way to explain the variation in the literature values at low redshift. The HB sample gives the lowest 
{\rm lum/faint} ratio of the three samples at all redshifts, which is easily explained as the selection algorithm for this sample is based 
on the cluster density profile fit, in contrast to the B and eBCS, which are based on the presence of bright red galaxies and BCGs; the 
close correlation between cluster X-ray brightness, used to select the eBCS, and BCG magnitude has been known for some time 
(e.g. \citealp{Edge-1991,Collins-1998}).

The degree of evolution in the {\rm lum/faint} ratio at 
high redshift is still somewhat confused. Measurement of this ratio has now been made in the highest redshift X-ray cluster known 
(J2215-1735) at $z=1.46$ \citep{Hilton-2009}. However, in contrast to the Andreon clusters at $z>1$ previously discussed, 
J2215 has a {\rm lum/faint} ratio of $2.2\pm0.9$ when transformed onto the De Lucia system, a value consistent with the prediction of 
$\simeq1.3$ based on a simple extrapolation of our best-fitting line in Fig. 4. 

The evidence for evolution in the {\rm lum/faint} ratio 
seen in Fig. 4 results from the deficit of faint galaxies on the red sequence in comparison to clusters observed at lower redshifts. Taken 
at face value, this is consistent with higher mass galaxies ending their star formation earlier than in their low mass counterparts; a 
process dubbed as ÔdownsizingÕ \citep{Cowie-1996}. However the question still remains as to the process by which the CMR 
becomes populated with RSGs and in particular whether the dominant mechanism is through merging or the stripping of spiral and irregular 
galaxies transforming them into passive S0s; an idea that is supported by the decrease in S0 galaxies along with the increasing fraction of 
spiral and irregular galaxies with redshift \citep{Dressler-1997, Smith-2005, Postman-2005}.

Our partial Spearman results offer at 
least one possible clue; the lack of an underlying correlation between {\rm lum/faint} ratio and $L_{\rm x}$ over our redshift range 
suggests that at least the late-time build up of the CMR is not related to processes associated with the hot intra cluster medium, such as 
ram pressure stripping or other mechanisms that depend on cluster mass, like tidal stripping or harassment 
\citep{Wake-2005, Mei-2009, Stott-2007}; however the large variation in the {\rm lum/faint} ratio for massive clusters at $z \geq 1$ 
previously mentioned, indicates that this issue is far from settled. Furthermore, our partial Spearman results highlight the importance of 
appropriate statistical analyses in determining the significance of possible correlation trends, particularly when faced with flux-limited 
samples. 

Turning to the possible role of mergers, since the fraction of massive early-type galaxies in clusters has been shown to 
remain consistently high out to $z=0.8$, the evolution seen in magnitude-limited samples may be dominated by fainter (sub-$M^{\ast}$ in the 
stellar mass) galaxies undergoing merging \citep{van-der-Wel-2007, Holden-2007}. An important recent development possibly related to this 
is the discovery of early-type massive compact ($\simeq 1$kpc) galaxies at $z\simeq2$ (e.g. \citealp{Trujillo-2006, Toft-2007, 
van-Dokkum-2009}). The dearth of such objects in local samples 
\citep{Taylor-2009} implies that these galaxies must undergo a rapid size evolution, growing by a factor 
$\simeq 4-5$ since $z=2-3$. Among the models that have been proposed the currently favoured mechanism driving this growth is also through 
minor merging with sub-$M^{\ast}$ galaxies (e.g. \citealp{Naab-2009}). Although most attention has focused on high redshift 
compact galaxies in the field, if the merging explanation is correct it should also apply to ellipticals in clusters. It therefore remains 
to be seen if a single sub-$M^{\ast}$ population in clusters can explain both the build up onto the CMR and the rapid size evolution of 
ellipticals. On the other hand, it needs to be stressed that the local dearth of these early-type massive compact galaxies and their
rapid size evolution are still a controversial issue. In fact, \citet{Valentinuzzi-2009} claimed that such objects might exist also locally 
in clusters, while \citet{Hopkins-2010} suggested different mechanisms driving a more modest size evolution. Finally, but not less 
important, \citet{Muzzin-2009} pointed out the estimated masses of these objects may be extremely uncertain.

We plan future investigations of the CMR using statistical samples of clusters over a wide redshift range from the serendipitous XCS 
cluster sample based on $XMM-Newton$ \citep{Sahlen-2009, Hilton-2009} whose flux limit 
($3.5\times 10^{-13}\ {\rm erg\ cm^{-2}\ s^{-1}}$) is an order of magnitude lower than the eBCS sample used here, which will enable trends 
with $L_{\rm X}$ to be reliably determined over a broad redshift range and allow us to investigate the CMR evolution in more detail in the 
redshift range $z=1-2$.\\

\section{Conclusions}
\label{sec:section8}
We study the {\rm lum/faint} ratio of RSGs for a large sample of optically (266) and X-ray (112) selected galaxy clusters in the sparsely 
covered regime ($0.05\leq z<0.19$) using data from the SDSS DR6 to  investigate how this ratio varies between different cluster samples 
(optical and X-ray) and to investigate possible trends with cluster mass and redshift, reported by other authors.

\begin{itemize}

\item[(i)] Independent of the method used, we find values of the {\rm lum/faint} ratio consistent with those 
found by \citet{De-Lucia-2007} for SDSS clusters, and a correlation with redshift [$\log({\rm lum/faint})=(1.2\pm 0.1) \log(1+z)$],
confirming a continuous trend in downsizing to low redshift.\\   

\item[(ii)] From a partial Spearman rank correlation test, we find no trend of {\rm lum/faint} ratio with $L_{\rm X}$ when correlations
between $L_{\rm X}$ and z and between {\rm lum/faint} ratio and z are removed, in agreement with the suggestion of 
\citet{Gilbank-Balogh-2008} and \citet{Andreon-2008}. This may be due to the narrow cluster mass range investigated.\\

\item[(iii)] The CMR slopes are $\sim -0.034$ for all the samples and consistent within $2\sigma$ of each other. These are similar to the values 
obtained in similar observational studies using similar rest-frame colours (e.g. \citealp{Stott-2009}); however they are inconsistent with 
the ones inferred through the theoretical model by \citet{Kodama-Arimoto-1997}. This may be due to the fact that this model is calibrated to the 
CMR of the Coma cluster, while we obtain slopes for stacked CMRs, containing several and possibly diverse clusters. 
\end{itemize}

\subsection*{Acknowledgments}
The authors thank the referee for his/her valid comments.\\
We thank Dr. G. De Lucia, for having kindly provided her data. We also thank Dr. T. Kodama for providing 
access to his models.\\
The authors acknowledge Dr. E. De Filippis for having provided the list of the control fields for the background subtraction and thank 
Dr. M. Hilton for helpful conversations.\\
DC expresses his gratitude to Dr. E. De Filippis, Dr. M. Paolillo, Prof. G. Longo and 
Dr. R. D'Abrusco, of the Department of Physical Sciences at the University of Napoli Federico II, for the valuable discussions. DC also 
thanks Dr. Cristiano Porciani for 
the useful suggestions.\\
Funding for the Sloan Digital Sky Survey (SDSS) and SDSS-II has been
provided by the Alfred P. Sloan Foundation, the Participating
Institutions, the National Science Foundation, the U.S. Department of
Energy, the National Aeronautics and Space Administration, the Japanese
Monbukagakusho, the Max Planck Society and the Higher Education
Funding Council for England. The SDSS Web site is http://www.sdss.org/.

\bibliographystyle{mn2e}

\bibliography{Reference}

\end{document}